\documentclass[
]{ceurart}

\usepackage[utf8]{inputenc} 
\usepackage[T1]{fontenc}    
\usepackage{hyperref}       
\usepackage{url}            
\usepackage{booktabs}       
\usepackage{amsfonts}       
\usepackage{nicefrac}       
\usepackage{microtype}      
\usepackage{xcolor}         
\usepackage{todonotes}
\usepackage{tcolorbox}
\usepackage{multirow} 
\usepackage{array}
\usepackage{threeparttable}
\usepackage{fancyvrb,fvextra} 
\usepackage{listings}

\begin{document}


\copyrightyear{2022}
\copyrightclause{Copyright for this paper by its authors.
  Use permitted under Creative Commons License Attribution 4.0
  International (CC BY 4.0).}
\conference{This is the author’s version of the work. It is posted here for your personal use.  The version of record is available in the proceedings of The first annual workshop on Learnersourcing: Student-generated Content @ Scale}


\title{Robosourcing Educational Resources -- Leveraging Large Language Models for Learnersourcing}


\author[1]{Paul Denny}[
orcid=0000-0002-5150-9806,
email=paul@cs.auckland.ac.nz,
]
\address[1]{The University of Auckland, Auckland, New Zealand}

\author[2]{Sami Sarsa}[
orcid=0000-0002-7277-9282,
email=sami.sarsa@aalto.fi,
]
\address[2]{Aalto University, Espoo, Finland}

\author[2]{Arto Hellas}[
orcid=0000-0001-6502-209X,
email=arto.hellas@aalto.fi,
]

\author[2]{Juho Leinonen}[
orcid=0000-0001-6829-9449,
email=juho.2.leinonen@aalto.fi,
]



%

\begin{abstract}
In this article, we introduce and evaluate the concept of robosourcing for creating educational content. Robosourcing lies in the intersection of crowdsourcing and large language models, where 
requests to large language models 
replace some of the work traditionally performed by the crowd.
Robosourcing includes a human-in-the-loop to provide priming (input) as well as to evaluate and potentially adjust the generated artefacts; these evaluations could also be used to improve the large language models. 
We explore the feasibility of robosourcing in the context of education by conducting an evaluation of robosourced programming exercises, generated using OpenAI Codex. Our results suggest that robosourcing could significantly reduce human effort in creating diverse educational content while maintaining quality similar to human-created content.  Thus, we argue that robosourcing has the potential to alleviate known issues around learner motivation and content quality that have been shown to limit the benefits of learnersourcing in practice.

\end{abstract}


\maketitle

\section{Introduction}

%
%
%
%
%
%
%


Learnersourcing is a broad term that is used to describe student-centered pedagogies that involve learners in the creation and evaluation of educational resources.  In contrast to more traditional models of teaching, where expert instructors assume the responsibility for producing the resources that are subsequently used by students, learnersourcing leverages the creativity and energy of a cohort of students in order to produce large repositories of useful learning content.  Khosravi et al. describe learnersourcing as a form of crowdsourcing that mobilizes students as ``experts-in-training to contribute to teaching or learning while being engaged in a meaningful learning experience themselves'' \cite{khosravi2021charting}.  

As highlighted by the previous definition, learnersourcing activities offer several benefits to students.  Broadly speaking, these benefits relate to the two primary activities underlying learnersourcing -- the generation of content, and the use of content generated by others.  When producing novel learning content, students must engage with and understand the concepts they are targeting.  The act of creating content leads to more robust recall of information when compared to passive engagement with content produced by others \cite{crutcher1989cognitive, dewinstanley2004processing}.  Generating model answers or solutions as part of this process prompts self-explanation which is also known to be beneficial to learning \cite{vanlehn1992model, chi1989self}.
When using content produced by other learners, there are benefits relating to the ways in which that content is presented and the quantity of resources available.  Learners appreciate the difficulties that their peers face and do not suffer from the phenomenon of expert blind spots which can make expert-generated resources challenging for novices to understand \cite{guo2020learnersourcing}.  Learnersourced repositories also scale with the size of the cohort, and thus provide a wide variety of content suitable for the needs of individual learners \cite{heffernan2016future, glassman2016learnersourcing}. 

However, despite the potential benefits offered by learnersourcing, in practice there are several challenges to successful implementation.  With respect to creating content, issues of \emph{low motivation} can prevent learners from properly engaging with the generative aspects of learnersourcing.  Indeed, prior research has shown that students tend to be much more inclined to use and evaluate resources created by others than they are to create resources themselves \cite{singh2021whats, pirttinen2022can, denny2018empirical}.  With respect to utilising the resources produced by other learners, one of the widely cited issues with learnersourcing is the \emph{low quality} of some of the content generated by novices \cite{yang2016student, abdi2021quality, purchase2010quality}.  Low quality resources are of limited use for learning.  Efforts to train novices to produce higher quality resources can help, but these can be time consuming and limit the scalability of learnersourcing in very large classes \cite{bates2014assessing}.

The recent emergence of large language models (LLMs) presents the possibility to scaffold learnersourcing activities in a way that may address, to some extent, the challenges relating to low student motivation and low quality content.  LLMs have proven to be remarkably adept at generating realistic human-like content of various types including text, images and source code.  Widely known models such as GPT-3~\cite{brown2020language}, OpenAI Codex~\cite{chen2021evaluating}, AlphaCode~\cite{li2022competition} and DALL$\cdot$E~\cite{ramesh2022hierarchical} have received a great deal of attention, and can produce novel content from a small number of contextual input examples \cite{lampinen2022language}.  The usage of large language models has grown dramatically in the last few years, following increases in the size of training data sets and the number of parameters used in the models.  They have been applied to a wide range of tasks and generated enormous quantities of new content, yet their potential impact on learners and on pedagogy remains to be fully explored.  

In this work, we propose `robosourcing' as a model to scaffold the creation of educational content.  Robosourcing combines large language models and learnersourcing by utilizing the large language models to facilitate the creation of content. A human (i.e. the learner) is involved in the robosourcing process as the initiator providing input to the models (e.g. content and topic priming) as well as evaluating and curating the content generated by the models.  Robosourcing shifts the primary focus of learners from content creation to content evaluation -- thus more accurately reflecting professional practice.  We evaluate preliminary evidence\footnote{Full evaluation to appear in \cite{sarsa2022automatic}.} highlighting the potential of robosourcing and argue that it may be an effective strategy for improving learner productivity and the quality of the content they generate when engaging in learnersourcing activities.


\section{Background}

\subsection{Practice testing}

A common type of learnersourcing task involves students creating practice questions which can then be used for drill and practice learning.  Across a wide range of educational contexts, drill and practice activities are both popular and highly effective \cite{dunlosky2013strengthening}.  The testing effect is a robust phenomenon which states that being tested on previously studied material is more effective for learning than repeated episodes of studying \cite{roediger2006test, Roediger_2013_RPT}.  This effect has been well established both in controlled laboratory environments and in the classroom, with clear evidence that frequent testing yields positive effects on both perceived and actual learning outcomes \cite{Bangert_1991_FCT}.

A primary challenge of supporting practice testing at scale is the human effort associated with generating questions and associated solutions.  Large repositories of questions that cover all relevant concepts are required in order to support effective practice \cite{ericsson2018cambridge}.  Given the critical role that feedback plays in learning ~\cite{hattie2007power, vollmeyer2005surprising}, the presence of model solutions is also important for providing immediate feedback to learners which can prompt reflection and promote self-regulated learning behaviours \cite{shute2008focus}.  Furthermore, research on the problem description effect has shown that the presence of relevant contextual information in the wording of a question can impact cognitive load and have a positive effect on problem success \cite{leinonen2021exploring, bouvier2016novice}.  The manual generation of suitable practice testing repositories, with comprehensive concept coverage, appropriate contextual information and model solutions, places a significant burden on instructors.

\subsection{Scalable problem generation}

A number of automated approaches have been explored for the generation of questions that could be used for practice testing.  These approaches commonly involve the use of question templates \cite{Zavala2018semantic, liu2018large} or parameterized questions \cite{Hsiao2009adaptive}.  In such cases, experts carefully design templates with selected elements that can be randomly generated or drawn from a pool of candidates.  Although such approaches can be used to generate a large number of questions, there remains significant human effort in constructing the templates and challenges in generating questions of equivalent difficulty when that is required \cite{denny2019fairness}.  

In certain domains, such as computer science and mathematics, specialized techniques can be employed to generate relevant problems.  For example, to generate code-tracing practice questions for computing students, Thomas et al. use a stochastic tree-based generation algorithm \cite{thomas2019stochastic}. This approach is effective at generating multiple-choice questions with good distractors, but the range of problems produced is very narrow.  Fowler and Zilles produce large pools of programming questions through a permutation strategy which makes superficial changes to the wording of the problem statement \cite{fowler2021superficial}.  Generating good base questions for permutation still requires manual human effort, and the number of permutations possible per base question is fixed.  In mathematics education, the use of word problems where numerical data is embedded in a natural language description of a scenario is ubiquitous \cite{verschaffel2020word}.  Many template-based approaches for generating such word problems have been explored, including manually defined templates with rules \cite{polozov2015personalized} and templates where nouns and verbs are replaced with appropriate words from a desired topic \cite{koncel2016theme}.  As with any template-based approach however, significant manual effort is required to generate new domain-specific templates of high quality.   

\subsection{Large language models}

Recently, there has been great progress on generative NLP methods (such as OpenAI's GPT-3~\cite{brown2020language}) that are capable of generating text that can be hard to distinguish from text written by humans~\cite{brown2020language}. These are deep learning models and their performance relies on both a vast number of parameters for the models (175 billion in the case of GPT-3) as well as an extensive corpus of text for training (570GB of text for GPT-3). Codex~\cite{chen2021evaluating}, also by OpenAI, is similar to GPT-3, but has been trained with a very large number of public source code repositories from GitHub. While GPT-3 is primarily used to create novel content based on existing content, the goal of Codex is to both translate natural language to source code and vice versa, and to generate/auto-complete source code from given source code. 

Both GPT-3 and Codex have been previously used to solve math and programming exercises respectively. For math word problems, Cobbe et al.~\cite{cobbe2021training} observed that while GPT-3 was relatively poor at solving primary school math problems, the performance could be improved by separately training verifiers that rank outputs from the large language model and using the best option as the solution. In the context of solving programming exercises, Finnie-Ansley et al.~\cite{finnie2022robots} observed that Codex was able to correctly answer most introductory programming problems and that when given typical introductory programming exam questions, Codex performed better than the average student.
Another LLM for code generation, AlphaCode~\cite{li2022competition} by DeepMind, was trained to solve competitive programming problems and managed roughly as well as an average competitive programmer on 10 Codeforces problems.

In this work, we explore the potential of such models for the scalable generation of practice questions as part of a learnersourcing activity.  Learners can leverage the computational effort of large language models to produce content quickly, which they can then evaluate and modify before sharing with others.  They can also provide basic priming information to the large language models, such as initial example problems and desired contextual information in the output.  We use the term robosourcing to describe this new approach.


\section{A Robosourcing model}

\begin{figure}[ht!]
\centering
\includegraphics[width=0.9\textwidth]{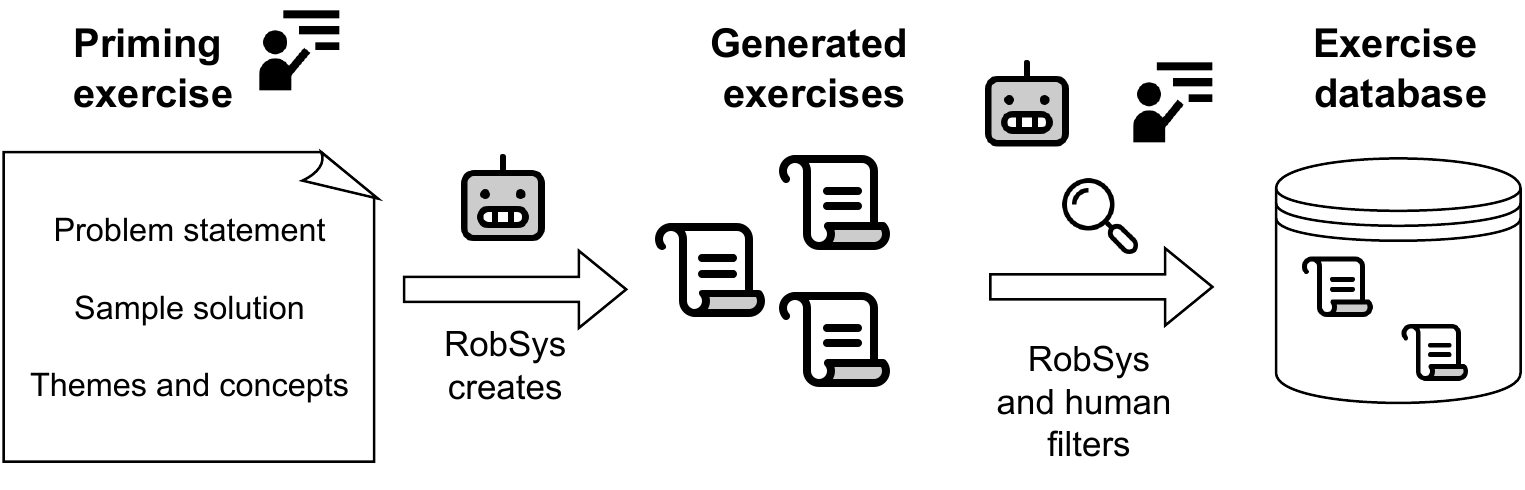}
\caption{Overview of the robosourcing model. A user (e.g. a learner) provides a starting point for creating exercises, which the system then uses to create an exercise pool. The exercise pool is filtered initially by the system based on options given by the user, after which the user can also filter out exercises. The exercises that are not filtered out are added to an exercise database.\label{fig:robsys}}
\end{figure}

A schematic view of the robosourcing model is shown in Figure~\ref{fig:robsys}; first, a learner provides a \emph{priming exercise} that includes a problem statement, a sample solution, and any expected themes and concepts (for programming exercises, automated tests can also be included). Once the priming exercise has been provided, the system automatically generates a pool of exercises for evaluation. This is followed by a filtering phase where the system first automatically filters generated exercises, after which the learner can perform additional filtering.  The filtered exercises can then be edited, if necessary, and added to an exercise database.  In a typical learnersourcing environment, students are involved in both creating and evaluating content.  This robosourcing model leverages the power of large language models for content creation, shifting the primary role of the learner towards evaluation.

In the first stage of the model, a learner provides a priming exercise as input to a large language model.  Consider the priming exercise for a programming problem, as shown below.  In this case, the format of the priming exercise consists of a label (i.e. Exercise 1) followed by keywords for both the contextual themes and the programming-related concepts within the exercise (i.e. donuts, function, conditional), a natural language problem statement and a solution (in the form of a Python function). For space reasons, we omit a list of test cases but these can also be included for programming problems.  The priming input ends with the explicit prompt for a new exercise to be generated (i.e. Exercise 2), along with the desired concepts and themes that it should target which are expressed as keywords (i.e. basketball, function, list and for loop). 

\begin{tcolorbox}
\begin{Verbatim}[fontsize=\fontsize{9pt}{9pt},breaklines]
"""Exercise 1
--Keywords--
donuts
function
conditional
--Problem statement--
Write a function called donut_cost that takes three inputs: the price of a donut, the number of donuts bought, and whether or not it is the weekend.  The function should return the total cost of the donuts.  If it is the weekend, the donuts cost twice as much.
--Sample solution--
def donut_cost(price, number, is_weekend):
  if is_weekend: 
    return price * number * 2
  else: 
    return price * number
"""Exercise 2
--Keywords--
basketball
function
list
for loop
--Problem statement--
\end{Verbatim}
\end{tcolorbox}

If this input is provided, verbatim, to the Davinci-2 model from OpenAI Codex~\cite{chen2021evaluating}, which is fine-tuned to generate code, one possible output is the following (note, in this case, the problem statement is related to basketball and the model solution consists of a function that involves a list and a for loop, as requested by the keyword information provided in the input):

\begin{tcolorbox}
\begin{Verbatim}[fontsize=\fontsize{9pt}{9pt},breaklines]
Write a function called count_rebounds that takes a list of basketball players as an input.  The function should return the total number of rebounds for the entire team.  Each element in the list is itself a list containing the player's name, their points, and their rebounds.
--Sample solution--
def count_rebounds(players):
  total = 0
  for player in players:
    total = total + player[2]
  return total
\end{Verbatim}
\end{tcolorbox}

In the second phase of the model, the learner then evaluates this output and makes a decision about whether it should be shared to a database of filtered problems, which can later be used for practice.  Some of this filtering can be performed automatically -- for example, in the case of programming problems, the generated code can be executed and tested against a suite of test cases (which can be also generated by the model, but were omitted for space reasons in this example).  Any generated problems for which the code cannot be executed, or for which the test cases do not pass, can also be automatically filtered out.   Finally, although in the example above we used the Davinci-2 model, the underlying generative model can be switched to something more appropriate for the problem type (such as GPT-3 or similar for contexts other than programming).  



\subsection{Preliminary evaluation}


To assess the feasibility of robosourcing, we use OpenAI Codex to generate a corpus of practice exercises suitable for university level introductory programming.  The goal of this evaluation is to determine whether the automatically generated exercises are of sufficient quality that they could be evaluated and modified with little effort by learners.  For more details on the evaluation, please see \cite{sarsa2022automatic}.

\subsubsection{Exercise generation}

To prime the exercises -- in other words, as input to the OpenAI Codex model -- we used a variant of the speeding problem presented in ~\cite{venables2009closer} and a currency converter program.  These two priming exercises are listed in Appendix~\ref{appendix:creation-priming-exercises}.  To explore the generation of varied and novel problems, we manipulate both the programming-related concepts and the contextual themes that are provided as part of the priming information as keywords.  We defined a total of nine contextual themes (e.g. hiking, music) and two distinct sets of programming-related concepts (e.g. set 2: class, list, conditional).  The full list of themes and concepts is provided in Table~\ref{tbl:priming-keywords}.  As shown in Appendix~\ref{appendix:creation-priming-exercises}, the input to the Codex model was the stop sequence (\texttt{"""}), followed by the label \texttt{Exercise 1} and then the complete priming exercise (the keywords, problem statement, sample solution and tests), followed by the stop sequence again.  The priming then continued with the label \texttt{Exercise 2}, the desired keywords selected from our themes and sets of programming concepts, and finally the label \texttt{Problem statement}.  The output from the model followed this label.  In this preliminary evaluation, we used the ``code-davinci-001'' model version of Codex.




\begin{table}[h!]
    \centering
    \caption{Keywords used for priming exercise generation. }
    \small
    \begin{tabular}{l|p{4.5cm}|p{4.5cm}}
    \toprule
    themes & programming concept set 1 & programming concept set 2  \\
    \midrule
    \multirow{5}{8em}{hiking,
        fishing,
        relationships,
        football,
        music,
        health,
        ice hockey,
        books,
        cooking
    }
    & function,   & class,       \\
    & parameters, & list,        \\
    & dictionary, &  conditional \\ 
    & arithmetic  &   \\
    &  &  \\
    \bottomrule
    \end{tabular}
    \label{tbl:priming-keywords}
\end{table}

We generated a total of 240 programming exercises.  These were a combination of the two programming exercises (see Appendix~\ref{appendix:creation-priming-exercises}), a total of nine themes (and an extra for leaving out the contextual concept) and two pro\-gram\-ming-related concept sets (and an extra for leaving out the programming-related concepts).  This resulted in $10 \times 3 \times 2 = 60$ different combinations of inputs (themes $\times$ programming-related concept sets $\times$ exercise primings). In addition, we explored two values for Codex's temperature parameter ($0$ and $0.75$) and created two exercises for each parameter combination. In total, this led to a sample of $60 \times 2 \times 2 = 240$ programming exercises.  We conducted the evaluation qualitatively and quantitatively.  For the qualitative evaluation, we assessed a random sample of 120 of these exercises. 


\subsubsection{Exercise evaluation}

For the qualitative part of the evaluation, we assessed the sensibleness, novelty, topicality, and readiness for use of the 120 randomly sampled exercises. For \emph{sensibleness}, we inspect whether the problem statement contains a sensible and practical problem that students might be expected to solve. For \emph{novelty}, we use Google search (on phrases contained within the problem statement) to see if we can find the exercise or a similar one online. For \emph{topicality} of the exercises, we analyse whether the generated problem incorporates the provided theme and concepts from the required sets.  

We also evaluate the \emph{readiness for use} of the exercises that were deemed to be sensible.  This category involved both a qualitative and a quantitative aspect.  For the qualitative piece, we evaluate whether the generated sample solution matches the problem description.  For the quantitative analysis, we examine the complete corpus of 240 programming exercises and assess three aspects.  We explore; 1) can the sample solution be executed/run, 2) does the sample solution pass the automated tests that are generated, and 3) what is the statement coverage of the automated tests when the code runs. These analyses were conducted programmatically\footnote{Analysis of statement coverage used \texttt{Coverage.py} version $6.3.2$ (https://coverage.readthedocs.io/)}.



The qualitative analysis was conducted by four researchers. Each researcher worked on a subsample of the exercises. The researchers assessed the items with \texttt{Yes} / \texttt{No} / \texttt{Maybe} statements, adding qualitative notes for the latter category. All the \texttt{Maybe} answers were then discussed by at least two researchers to determine the consensus label (\texttt{Yes} or \texttt{No}).  We tallied the \texttt{Yes} / \texttt{No} answers to provide quantitative results.


\section{Preliminary Findings}

The statistics for sensibleness, novelty, readiness for use and topicality of the evaluated programming exercises are presented in Table~\ref{tab:programming-exercise-results}. Out of the evaluated programming exercises, 75.0\% were sensible, 81.8\% were novel, and 76.7\% had a matching sample solution. In addition, 79.2\% of the exercises matched the priming theme, and more than three-quarters had the desired programming concepts present. The most common reason for a programming exercise not being sensible was the problem statement asking for the calculation of some sort of a value that required another value that was not available (e.g. amount of tax but missing price). In terms of the programming concepts, `arithmetic' was missing relatively often despite the programs featuring a `+'-symbol; in these cases, `+' was used for concatenating strings. 

 


\newcolumntype{x}[1]{>{\centering\let\newline\\\arraybackslash\hspace{0pt}}p{#1}}
\begin{table}[ht!]
    \centering
    \small
    \caption{Summary of the manually evaluated programming exercises.}
    \begin{tabular}{ c c c x{1.5cm} x{1.5cm} x{2.5cm} x{2.5cm} }
         \toprule
         Exercises & Sensible & Novel & Matches sample solution & Matches priming theme & Matches priming concepts (function/class) & Matches priming concepts (list/dictionary)\\
         \midrule
         120 & 75.0\% & 81.8\% & 76.7\% & 79.2\% & 78.3\% & 75.8\% \\
         \bottomrule
    \end{tabular}
    \label{tab:programming-exercise-results}
\end{table}

The statistics for the programmatic analysis of all 240 generated exercises are presented in Table~\ref{tab:dynamic_analysis_results}. Out of the 240 programming exercises, 203 had a sample solution (84.6\%). From the 203 sample solutions, 182 (89.7\%) could be executed (i.e.\@ running the code did not produce any errors). A total of 170 programming exercises had automated tests, while 165 programming exercises had both a sample solution and automated tests. From these 165 programming exercises, 51 had a sample solution that passed the automated tests. Out of the 51 programming exercises with a working sample solution and automated tests, 48 exhibited 100\% statement coverage.  The statement coverage averaged over all of the 51 programming exercises was 98.0\%.  We observed that the most common issue preventing exercises from passing the tests was not related to the code logic, but in how the outputs were handled. In those cases, the sample solution printed a value, while the automated tests expected that the sample solution would return a value (e.g.\@ the tests called a function and expected that the function would return a value, but the function printed a value). 
We note, of course, that a confusion between printing and returning values is a commonly cited error made by novices \cite{ettles2018common, izu2018can}.
Another common issue was that the tests expected specific numbers that were not possible with the inputs (e.g.\@ checking whether a program correctly extracted and returned a list of even values from a list received as a parameter, a test provided the list \texttt{[1, 2, 3]} as an input to the function and expected that the function would return the list \texttt{[2, 4]}).

Overall, most generated exercises satisfied most of the criteria.  We noted that in multiple cases, only minor tweaks would have been necessary to transform failing tests into passing ones. In the cases where tests were missing, we could simply add the generated exercise to the initial priming and the tests would likely be generated on a ``second'' run (we tested this behavior directly when exploring the output).  The results of this preliminary evaluation provide some support for our hypothesis that these automatically generated exercises could serve as useful starting materials for learners within a learnersourcing activity.  Learners could quickly evaluate the potential of a given exercise, and generate new ones if necessary.  With reasonably little effort, and certainly with less effort than producing an exercise from scratch, such exercises could be edited or modified and then contributed to a shared learning resource.

\begin{table*}[ht!]
\centering
\small
\caption{Summary of programmatic analysis of generated programming exercises \label{tab:dynamic_analysis_results}}
    \begin{tabular}{c|cccccc}
         \toprule
                     & Has solution?  & Solution runnable?  & Has tests?  & All tests pass?  & Test coverage \\
        \midrule
         \% & 84.6\%                & 89.7\%                            & 70.8\%      & 30.9\%           & 98.0\%       \\
         n / N  & 203 / 240             & 182 / 203                         & 170\footnotemark[1] / 240       & 51 / 165\footnotemark[1]            & 48\footnotemark[2] / 51  \\
         \bottomrule
        \multicolumn{6}{c}{
        \multirow{1}{13.6cm}{
            \footnotesize{$^1$Five of the exercises contained -{}-Tests-{}- but not -{}-Sample solution-{}- (needed for content extraction)}
        }
        }\\
        \multicolumn{6}{l}{
        \multirow{2}{13.6cm}{
            \footnotesize{$^2$The n out of N for test coverage is counted as the number of full coverage (100\%) cases out of the number of all test suites that did not fail (i.e.\@ when coverage can be computed)}
        }
        }
        
    \end{tabular}
\end{table*}

Below, we show a verbatim example of a well formed programming exercise generated by OpenAI Codex that received a positive score on 
our evaluation criteria.  It is generated using the ``speeding'' priming example exercise (in Appendix~\ref{appendix:creation-priming-exercises}), theme ``music'', and concept set of ``class'', ``list'' and ``conditional''.

\begin{tcolorbox}
\begin{Verbatim}[fontsize=\fontsize{9pt}{9pt},breaklines]
Assume that you are building a music library app. Create a class called MusicLibrary.
This class should have a constructor that takes in a list of dictionaries,
each of which has keys "name" and "artist". The constructor should create a
property called library that stores the list of dictionaries.
The class must also have a method called search that takes in the name of a song,
and returns the artist who sings it. If the song does not exist in the library,
the method should return None. The classes should also have a method called
get_artist_songs that takes in an artist name, and returns a list of songs from
the library by that artist.
--Sample solution--
class MusicLibrary:
  def __init__(self, songs):
    self.library = songs
  def search(self, name):
    for song in self.library:
      if song['name'] == name:
        return song['artist']
    return None
  def get_artist_songs(self, artist):
    songs = []
    for song in self.library:
      if song['artist'] == artist:
        songs.append(song['name'])
    return songs
--Tests--
class Test(unittest.TestCase):
  def test_library(self):
    songs = [{'name': 'Rude', 'artist': 'Magician'}, {'name': 'Sugar', 'artist': 'Maroon 5'}]
    library = MusicLibrary(songs)
    self.assertEquals(library.search('Rude'), 'Magician')
    self.assertEquals(library.search('Sugar'), 'Maroon 5')
    self.assertEquals(library.search('Love Yourself'), None)
    self.assertEquals(library.get_artist_songs('Magician'), ['Rude'])
    self.assertEquals(library.get_artist_songs('Maroon 5'), ['Sugar'])
\end{Verbatim}
\end{tcolorbox}


\section{Conclusions}


These preliminary results demonstrate that robosourcing is indeed a feasible approach for the scalable generation of practice questions as part of a learnersourcing activity.  The computational power of large language models (LLMs) can be utilised to generate artefacts that are of sufficient quality that they can be reviewed quickly, thus shifting the focus of learners from content creation towards content evaluation.  Robosourcing may help to address the issues relating to low learner motivation and low quality content which  are known to negatively affect learnersourcing activities in practice. 

We observed that approximately one third of the generated programming exercises would be directly usable for teaching, and certainly suitable as a starting point for learners to evaluate and modify.  The robosourced exercises were predominantly novel as they could not be found via Google search, and the themes and domain-specific concepts can be easily adjusted to influence the output.  At least from this initial exploration, LLMs such as OpenAI Codex appear to be useful tools for robosourcing a diverse and large pool of exercises.
The possibility for learners themselves to influence and customize the content through appropriate choice of keywords may also increase motivation.  Exploring student perceptions of the use of such models as part of a robosourcing activity would be a fascinating avenue for future work. 




 



Although these results are promising, there is still need for caution. 
Prior work on LLMs has found that they can contain biases, for example, towards minorities~\cite{solaiman2021process,xu2021detoxifying}. Indeed, we did observe that the exercises generated during our evaluation did seem to more commonly involve men in problem descriptions. However, we did not observe any offensive content. A similar concern is that LLMs might leak personally identifiable information present in the training data~\cite{carlini2021extracting}, although we did not observe any such data in our evaluation.  In addition, recent work in the programming education literature has found that LLMs can be used to solve introductory programming exercises~\cite{finnie2022robots}. Thus, familiarizing students with these technologies could increase their use for plagiarism -- and it is potentially even more likely that exercises created by LLMs can be solved by LLMs compared to exercises created by teachers.

Overall, while our results suggest robosourcing is feasible already, the performance of the underlying generative models is almost certain to improve over time, and will thus likely also lead to improved performance when these models are used for robosourcing.  For example, newer LLMs such as Gopher~\cite{rae2021scaling} by DeepMind outperform GPT-3 (which Codex is based on). Similarly, we used Codex without any fine-tuning: future work should explore fine-tuning LLMs specifically for robosourcing, which is likely to lead to further improvements in performance. Lastly, the LLMs used in robosourcing could themselves be improved by feedback from human-in-the-loop evaluations, increasing their performance for robosourcing tasks.

\bibliography{99-icer-refs.bib,99-lessons-learned.bib}

\newpage
\section{Appendix}

\subsection{Priming exercises for programming problems \label{appendix:creation-priming-exercises}}

The following two snippets were used as the priming exercises for robosourcing programming exercises. The keywords encompass the themes and concepts, which are followed by the problem statement and the sample solution. The place marked with `(themes and concepts are entered here)' is filled by the robosourcing system, while the used large language model generates content starting at `(generation starts here)'.  

\begin{tcolorbox}
\begin{Verbatim}[fontsize=\fontsize{9pt}{9pt},breaklines]
"""Exercise 1
--Keywords--
cars
function
parameters
conditional
--Problem statement--
Write a function called speeding_check that takes a single
parameter speed and prints out "You are fined for $200" if the speed is above 120,
"You are fined for $100" if the speed is above 100 but below 120
and otherwise prints "All good, race ahead".
--Sample solution--
def speeding_check(speed):
  if speed > 120:
    return "You are fined for $200"
  elif speed > 100:
    return "You are fined for $100"
  else:
    return "All good, race ahead"
--Tests--
class Test(unittest.TestCase):
  def test_speeding_check(self):
    self.assertEquals(speeding_check(100), 'All good, race ahead')
    self.assertEquals(speeding_check(101), 'You are fined for $100')
    self.assertEquals(speeding_check(121), 'You are fined for $200')
"""Exercise 2
--Keywords--
(themes and concepts are entered here)
--Problem statement--
(generation starts here)
\end{Verbatim}
\end{tcolorbox}

\begin{tcolorbox}
\begin{Verbatim}[fontsize=\fontsize{9pt}{9pt},breaklines]
"""Exercise 1
--Keywords--
currency
class
function
parameters
dictionary
arithmetics
--Problem statement--
Write a class called Converter that is initialized with
a dictionary of exchange rates for currencies against the USD, e.g. {"USD": 1, "EUR": 0.9, "GBP": 0.75}.
The class should have a method called convert, which takes in three parameters:
from_currency, to_currency, and amount.
The function should return the given amount converted from the first currency (first parameter)
to the second currency (second parameter) using the exchange rate dictionary given in the class constructor.

As an example, the code
converter = Converter({"USD": 1, "EUR": 0.9, "GBP": 0.75})
in_euros = converter.convert("GBP", "EUR", 10)
print(in_euros)
should print out 12.0
--Sample solution--
class Converter():
  def __init__(self, exchange_rates):
    self.exchange_rates = exchange_rates

  def convert(self, from_currency, to_currency, amount):
    amount_in_usd = amount / self.exchange_rates[from_currency]
    return amount_in_usd * self.exchange_rates[to_currency]
--Tests--
class TestConverter(unittest.TestCase):
  def test_converter(self):
    converter = Converter({"USD": 1, "EUR": 0.8})
    self.assertEquals(converter.convert("USD", "EUR", 100), 80)

  def test_converter2(self):
    converter = Converter({"USD": 1, "EUR": 0.9, "GBP": 0.75, "SEK": 9.71})
    self.assertEquals(converter.convert("USD", "USD", 100), 100)
    self.assertEquals(converter.convert("USD", "EUR", 100), 90)
    self.assertEquals(converter.convert("GBP", "EUR", 10), 12)
    self.assertEquals(converter.convert("EUR", "GBP", 10), 8.333333333333332)
"""Exercise 2
--Keywords--
(themes and concepts are entered here)
--Problem statement--
(generation starts here)
\end{Verbatim}
\end{tcolorbox}


\end{document}